\documentclass[aps,pra,showpacs,twocolumn,superscriptaddress]{revtex4}

\usepackage{latexsym}
\usepackage{graphics}
\usepackage{amsmath}
\usepackage{bm}

\begin{document}

\title{Probabilistic quantum multimeters}

\author{Jarom\'{\i}r Fiur\'{a}\v{s}ek}

\affiliation{Ecole Polytechnique, CP 165, Universit\'{e}
     Libre de Bruxelles, 1050 Bruxelles, Belgium}
\affiliation{Department of Optics, Palack\'y University,
     17.~listopadu 50, 772\,00 Olomouc, Czech~Republic}

\author{Miloslav Du\v{s}ek}

\affiliation{Department of Optics, Palack\'y University,
     17.~listopadu 50, 772\,00 Olomouc, Czech~Republic}

\begin{abstract}
We propose quantum devices that can realize probabilistically
different projective measurements on a qubit. The desired
measurement basis is selected by the quantum state of a program
register. First we analyze the phase-covariant multimeters for a
large class of program states, then the universal multimeters for
a special choice of program. In both cases we start with
deterministic but erroneous devices and then proceed to devices
that never make a mistake but from time to time they give an
inconclusive result. These multimeters are optimized (for a given
type of a program) with respect to the minimum probability of
inconclusive result. This concept is further generalized to the
multimeters that minimize the error rate for a given probability
of an inconclusive result (or vice versa). Finally, we propose a
generalization for qudits.
\end{abstract}


\maketitle

\section{Introduction}

\begin{figure}
\centerline{\resizebox{0.9\hsize}{!}{\includegraphics*{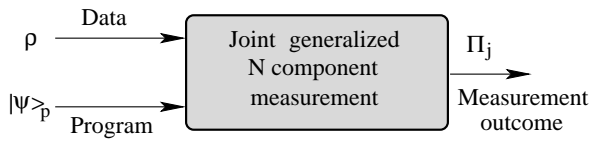}}}
\caption{Schematic drawing of a quantum multimeter. The effective
measurement performed on the data state $\rho$ is selected by the
quantum state of the program register $|\Psi\rangle_{\mathrm{p}}$.
The multimeter itself carries out a \emph{fixed} joint generalized
measurement on data and program states which is described by a
POVM $\{\Pi_j\}$.} \label{fig0}
\end{figure}

Programmable quantum multimeters are devices that can realize any
desired generalized quantum measurement from a chosen set (either
exactly or approximately) \cite{DuBu,FiDuFi}. Their main feature
is that the particular positive operator valued measure (POVM) is
selected by the quantum state of a ``program register'' (quantum
software). In this sense they are analogous to universal quantum
processors \cite{Nielsen97,Vidal00,Hillery02,Hillery02b}. The multimeter
itself is represented by a \emph{fixed} joint POVM on the data and
program systems together (see Fig.~\ref{fig0}). Each outcome of
this POVM is associated with one outcome of the ``programmed''
POVM on the data alone. From the mathematical point of view the
realization of a particular quantum multimeter is equivalent to
the optimal discrimination of certain mixed states. A different
kind of a quantum multimeter that can be programmed to evaluate
the expectation value of any operator has been introduced in
Ref.~\cite{Paz03}. Besides quantum multimeters, other devices whose
operation is based on the joint measurement on two different
registers have been proposed recently. The universal quantum
matching machine that allows to decide which template state is
closest to the input feature state was analyzed in \cite{Sasaki02}.
The problem of comparison of quantum states was studied in \cite{BaChJe}.
The so-called universal quantum detectors have been considered
in \cite{DAriano03}. All these devices could play an important
role in quantum state estimation and quantum information
processing.

In this paper, we will describe programmable quantum devices
that can accomplish von Neumann measurements on a single qubit.
However, it is impossible to perfectly encode arbitrary projective measurement
on a qubit into a state in finite-dimensional Hilbert space \cite{DuBu}.
The proof of this theorem is similar to the proof
that it is impossible to encode an arbitrary unitary operation
(acting on a finite-dimensional Hilbert space) into a state of a
finite-dimensional quantum system \cite{Nielsen97}. Briefly, one can show that
any two program states that perfectly encode two different measurement
bases must be mutually orthogonal. Nevertheless, it is still possible to
encode POVMs that represent, in a certain sense, the best
approximation of the required projective measurements.

A specific way of approximation of projective measurements is a
``probabilistic'' measurement that allows for some inconclusive
results. In this case, instead of a two-component projective measurement
one has a three-component POVM and the third outcome corresponds to
the inconclusive result. The natural request is to minimize the
error rate at the first two outcomes. As a limit case it is
possible to get an error-free operation (however, with a nonzero
probability of an inconclusive result) -- such a multimeter
performs the exact projective measurements but with the
probability of success lower than one. Such a device is conceptually
analogous to the probabilistic programmable quantum gates
\cite{Nielsen97,Vidal00,Hillery02}. The other boundary case is an
ambiguous multimeter without inconclusive results \cite{FiDuFi}.

Our present article is organized as follows. In Sec.~\ref{PCM} we
start with the analysis of phase-covariant multimeters
 that can perform von Neumann measurement on a single qubit
in any basis located on the equator of the Bloch sphere. First we
discuss deterministic devices (no inconclusive results but errors
may appear), then error-free probabilistic devices (no errors but
inconclusive results may appear), and finally general multimeters
with given fraction of inconclusive results optimized with respect
to minimal error-rate. In this section we also introduce and
explain in detail all necessary mathematical tools. Further, in
Sec.~\ref{UM} we study universal multimeters that can accomplish
\emph{any} von Neumann measurement on a single qubit. We
confine our investigation to the program consisting of the two
basis vectors. Again, we start with deterministic devices,
continue with error-free multimeters and finally proceed to
apparatuses with a given fraction of inconclusive results.
Sec.~\ref{qudits} is devoted to probabilistic error-free universal
multimeters that can accomplish \emph{any} projective measurement
on a \emph{qudit}. Sec.~\ref{concl} concludes the paper with a
short summary.

\section{Phase-covariant multimeters\label{PCM}}

In this section we will consider multimeters that should perform
von Neumann measurement on a single qubit in any basis
$\{|\psi_{+}\rangle,|\psi_{-}\rangle\}$ located on the equator of
the Bloch sphere,
\begin{equation}
|\psi_{\pm}(\phi)\rangle=\frac{1}{\sqrt{2}}(|0\rangle\pm e^{i\phi}|1\rangle),
\label{pcbasis}
\end{equation}
where $\phi\in[0,2\pi]$ is arbitrary. To simplify notation, we
shall not usually display the dependence of the basis states on
$\phi$ explicitly in what follows. Generally, the design of the
optimal multimeter should involve the optimization of both the
dependence of the program on the measurement basis and the fixed
joint measurement on the program and data states. However, this is
a very hard problem that we will not attempt to solve in its
generality. Instead, we will design an optimal multimeter for a
particular simple and natural choice of the program. Namely,
similarly as in \cite{FiDuFi}, we assume that the program of the
multimeter $|\Psi\rangle_{\mathrm{p}}$ which determines the
measurement basis consists of $N$ copies of the basis state
$|\psi_{+}\rangle$,
$|\Psi\rangle_{\mathrm{p}}=|\psi_+\rangle^{\otimes N}$. Since we
have restricted ourselves to the bases (\ref{pcbasis}), the state
$|\psi_{-}\rangle$ can be obtained form $|\psi_{+}\rangle$ via
unitary transformation,
\begin{equation}
|\psi_{-}\rangle=\sigma_z|\psi_{+}\rangle,
\end{equation}
where $\sigma_z$ denotes the Pauli matrix. This implies that
all the programs of the form $|\psi_{+}\rangle^{\otimes j}
|\psi_{-}\rangle^{\otimes N-j}$ are equivalent to the program
$|\psi_{+}\rangle^{N}$ because these  programs are related via
a \emph{fixed} unitary $U= \openone^{\otimes j}\otimes \sigma_z^{\otimes N-j}$.
First, we shall derive the optimal deterministic multimeter, which
always yields an outcome, but errors may occur. Then, we shall consider
a probabilistic multimeter that conditionally  realizes exactly the
von-Neumann measurement in basis (\ref{pcbasis}), but at the expense
of some fraction of inconclusive results. Finally, we will show that
the deterministic and unambiguous multimeters are just two extremal cases
from a whole class of optimal multimeters that are designed such that the
probability of correct measurement on basis states is maximized for a fixed
fraction of inconclusive results.

\subsection{Deterministic multimeter\label{PCM-D}}

The multimeter is a device that performs a joint generalized
measurement described by the POVM $\{\Pi_j\}$ on the data state
and the program state, see Fig.~\ref{fig0}. This \emph{fixed}
joint measurement on the data and program can be also interpreted
as an effective measurement on the data register, which is
described by the POVM ${\pi_j}$ and depends of the program via
\begin{equation}
\pi_{j}= \mathrm{Tr}_{\mathrm{p}}[(\openone_{\mathrm{d}} \otimes |\Psi\rangle_{\mathrm{p}}\langle
\Psi|)\, \Pi_j],
\label{pcdataPOVM}
\end{equation}
where the subscripts $\mathrm{d}$ and $\mathrm{p}$ denote the data and program
states, respectively. The deterministic single-qubit multimeter is
fully characterized by a two-component POVM $\{\Pi_{+},\Pi_{-}\}$.
The readout of $\Pi_{+}$ is interpreted as the finding of the data
state in basis state $|\psi_{+}\rangle$ while $\Pi_{-}$ is
associated with the detection of $|\psi_{-}\rangle$. Ideally,
\begin{equation}
\pi_{\pm}=|\psi_{\pm}\rangle\langle \psi_{\pm}|
\label{pcdataPOVMideal}
\end{equation}
should hold, but this cannot be achieved for all $\phi$
with a finite-dimensional program.

The performance of the multimeter is quantified by the probability
$P_S$ that the measurement yields correct outcome when the data
register is prepared in the basis state $|\psi_{+}\rangle$ or
$|\psi_{-}\rangle$ with probability $1/2$ each. For each
particular phase $\phi$ we thus have
\begin{eqnarray}
P_S(\phi)&=&\frac{1}{2}\mathrm{Tr}
[\Pi_{+} \psi_{+}(\phi)\otimes\psi_{+}^{\otimes N}(\phi)]+
\nonumber \\
& & \frac{1}{2}\mathrm{Tr}[\Pi_{-}
\psi_{-}(\phi)\otimes\psi_{+}^{\otimes N}(\phi)],
\end{eqnarray}
where $\psi_{\pm}=|\psi_{\pm} \rangle\langle \psi_{\pm}|$.
Assuming homogeneous a-priori distribution of the angle $\phi$ we
define the average success rate as
\begin{equation}
P_S=\int_0^{2\pi} P_S(\phi) \frac{d\phi}{2\pi}.
\label{pcPSdefine}
\end{equation}
We define the optimal deterministic multimeter as the
multimeter that maximizes $P_S$ for the program
$|\psi_{+}\rangle^{\otimes N}$. The choice of $P_S$ as the figure of merit is
strongly supported by the observation that $P_{S}$ can be interpreted as the
average \emph{fidelity} of the multimeter. Consider the effective POVM on the
data qubit $\{\pi_+(\phi), \pi_{-}(\phi)\}$ for some particular phase $\phi$.
It is natural to define the fidelity of this POVM with respect to the
projective measurement in the basis $|\psi_{\pm}(\phi)\rangle$ as follows,
\begin{eqnarray*}
F(\phi)&=&\frac{1}{2}\langle\psi_+(\phi)|\pi_{+}(\phi)|\psi_{+}(\phi)\rangle +
\nonumber \\
&&\frac{1}{2}\langle\psi_{-}(\phi)|\pi_{-}(\phi)|\psi_{-}(\phi)\rangle.
\end{eqnarray*}
It is easy to see that the average fidelity
$F=\frac{1}{2\pi}\int_{0}^{2\pi}F(\phi) d\phi$ coincides with the
average success rate (\ref{pcPSdefine}). Clearly, $F\leq 1$
and $F=1$ if and only if (\ref{pcdataPOVMideal}) holds for all
$\phi$ (maybe except of a set of measure zero).

To simplify the notation we introduce the symbol $C_{N,k}$ for the
binomial coefficient,
\begin{equation}
C_{N,k}={N \choose k}.
\end{equation}
On inserting the formula for $P_S(\phi)$ into
Eq.~(\ref{pcPSdefine}) and carrying out the integration over
$\phi$ we find that
\begin{equation}
P_S=\frac{1}{2}(\mathrm{Tr}[\Pi_{+} R_{+}]+ \mathrm{Tr}[\Pi_{-} R_{-}]),
\label{pcPS}
\end{equation}
where the two positive semidefinite operators $R_{\pm}$ read
\begin{eqnarray*}
R_{+}&=&\frac{1}{2^{N+1}}\sum_{k=1}^N C_{N+1,k}|
\varphi_{N,k}^{+}\rangle\langle \varphi_{N,k}^{+}| +\frac{1}{2^{N+1}}X
\nonumber \\
R_{-}&=&\frac{1}{2^{N+1}}\sum_{k=1}^N C_{N+1,k}
|\varphi_{N,k}^{-}\rangle\langle \varphi_{N,k}^{-}| +\frac{1}{2^{N+1}} X.
\end{eqnarray*}
Here
\begin{equation}
|\varphi_{N,k}^{\pm}\rangle=
\sqrt{1-B_{N,k}}\,|0\rangle_{\mathrm{d}}|N,k\rangle_{\mathrm{p}}
\pm
\sqrt{B_{N,k}}\,|1\rangle_{\mathrm{d}}|N,k-1\rangle_{\mathrm{p}},
\label{pcvarphiNk}
\end{equation}
with $B_{N,k}=k/(N+1)$. The operator $X$ that is common to
$R_{+}$ and $R_{-}$ is given by
\begin{equation}
X=|0\rangle_{\mathrm{d}}\langle 0| \otimes |N,0\rangle_{\mathrm{p}}\langle N,0| +
|1\rangle_{\mathrm{d}}\langle 1|\otimes |N,N\rangle_{\mathrm{p}}\langle N,N|,
\end{equation}
and $|N,k\rangle$ denotes a normalized totally symmetric state of
$N$ qubits with $k$ qubits in state $|1\rangle$ and $N-k$ qubits in
state $|0\rangle$.

It follows from Eq.~(\ref{pcPS}) that the optimal deterministic
multimeter is the one that optimally discriminates between two
mixed states $R_{+}$ and $R_{-}$.  This problem has been analyzed
by Helstrom \cite{helst} who showed that the maximal achievable success rate
is
\begin{equation}
P_{S,\mathrm{max}}= \frac{1}{2} + \frac{1}{4}\mathrm{Tr}|R_{+}-R_{-}|
\end{equation}
and the optimal POVM is given by projectors onto the subspaces spanned
by the eigenstates of $\Delta R=R_{+}-R_{-}$ with positive and negative
eigenvalues, respectively. If some of the eigenvalues of $\Delta R$ are
zero, then the projectors can be freely added either to $\Pi_{+}$
or $\Pi_{-}$.

In the basis $|0\rangle_{\mathrm{d}}|N,k\rangle_{\mathrm{p}}$, $|1\rangle_{\mathrm{d}}|N,k\rangle_{\mathrm{p}}$,
the matrix $\Delta R$ is block diagonal and its eigenvalues and
eigenstates can easily be determined.  Since $\mathrm{Tr}|\Delta R|$
is equal to the sum of absolute values of the eigenvalues of $\Delta R$,
we find after simple algebra that
\begin{equation}
P_{S,\mathrm{max}}=\frac{1}{2}+ \frac{1}{2^{N+1}}\sum_{k=1}^N
\sqrt{{ N \choose k}{N \choose k-1}}.
\label{pcPSmaxdeterministic}
\end{equation}

Interestingly enough, $P_{S,\mathrm{max}}$ is equal to the optimal
fidelity of estimation of $|\psi_{+}(\phi)\rangle$ from $N$ copies
of $|\psi_{+}(\phi)\rangle$ \cite{DeBuEk}. So one possible
implementation of the optimal deterministic phase-covariant
multimeter with program $|\psi_{+}(\phi)\rangle^{\otimes N}$ would be to
first carry out the optimal estimation of $|\psi_{+}(\phi)\rangle$
and then measure the data qubit in the basis spanned by the
estimated state and its orthogonal counterpart. Instead, one could
also perform a joint generalized measurement on data and program.
The two POVM elements are given by
\begin{equation}
\Pi_{\pm}=\sum_{k=1}^N |\Pi_{N,k}^{\pm}\rangle\langle \Pi_{N,k}^{\pm}|
+\frac{1}{2} X,
\end{equation}
where
\begin{equation}
|\Pi_{N,k}^{\pm}\rangle=\frac{1}{\sqrt{2}}(|0\rangle_{\mathrm{d}}|N,k\rangle_{\mathrm{p}}
\pm |1\rangle_{\mathrm{d}}|N,k-1\rangle_{\mathrm{p}} ).
\end{equation}
The effective POVM on the data register (\ref{pcdataPOVM}) can be expressed
as
\begin{equation}
\pi_{\pm}=P_{S,\mathrm{max}}|\psi_{\pm}\rangle\langle \psi_{\pm}|+
(1-P_{S,\mathrm{max}})
|\psi_{\mp}\rangle\langle \psi_{\mp}|.
\label{pcdataPOVMoptimal}
\end{equation}
In the limit of infinitely large program register, $N\rightarrow \infty$, the
POVM (\ref{pcdataPOVMoptimal}) approaches the ideal projective measurement
(\ref{pcdataPOVMideal}).

\subsection{Error-free probabilistic multimeter\label{PCM-EF}}

The multimeter designed in the preceding section is only approximate, because
the effective POVM  (\ref{pcdataPOVMoptimal}) on the data register
differs from the projective measurement in the basis $|\psi_{+}\rangle$,
$|\psi_{-}\rangle$. Here, we construct a multimeter that
realizes  an \emph{exact} von Neumann measurement in the basis (\ref{pcbasis})
with some probability $P_S$. This is achieved at the expense of the
inconclusive results which occur with the probability $P_{I}=1-P_S$ and are
associated with the POVM element $\Pi_{?}$. Such a \emph{probabilistic}
multimeter must unambiguously discriminate between two mixed states
$R_{+}$ and $R_{-}$. The unambiguous discrimination of mixed quantum
states \cite{RuSpTu,RaLuEn}
(and, more generally, discrimination of mixed states with
inconclusive results \cite{FiJe,El}) has attracted a considerable
attention recently.

As formally stated in Ref.~\cite{RuSpTu},  we have to find a three-component POVM
$\Pi_+,\Pi_-,\Pi_?$  that maximizes the success rate (\ref{pcPS})
under the constraints
\begin{eqnarray}
\mathrm{Tr}[\Pi_{+}R_{-}] = \mathrm{Tr}[\Pi_{-}R_{+}]&=&0,
\nonumber \\
\Pi_{+}+\Pi_{-}+\Pi_{?}&=&\openone, \nonumber \\
\Pi_{+}\geq 0, \quad \Pi_{-}\geq 0, \quad \Pi_{?}&\geq& 0,
\label{pcconstraints}
\end{eqnarray}
which is an instance of the so-called semidefinite program.
The first constraint guarantees that the multimeter will never
respond with a wrong outcome, i.e.  $\Pi_{-}$ ($\Pi_{+}$)
cannot be detected when the data register is in the basis state
$|\psi_{+}\rangle$ ($|\psi_{-}\rangle$). The second and third
constraints express the completeness of the POVM and the  positive
semidefiniteness of the POVM elements.

Here we shall give a simple intuitive construction of the optimal POVM
and we shall analyze the dependence of $P_I$ on $N$.
The optimality of the POVM will be formally proved in the next
subsection using the techniques introduced in Ref.~\cite{FiJe}.

Due to the particular structure of the operators $R_{+}$ and $R_{-}$
the problem of unambiguous discrimination of $R_{+}$ and $R_{-}$
splits into $N$ independent problems of unambiguous discrimination
of two \emph{pure} states $|\varphi_{N,k}^{+}\rangle$ and
$|\varphi_{N,k}^{-}\rangle$. The unambiguous discrimination of two pure
non-orthogonal states with equal a-priori probabilities has been studied
by Ivanovic \cite{ivanovic}, Dieks \cite{dieks}, and Peres \cite{peres}
(IDP). The minimal probability of
inconclusive results is equal to the absolute value of the scalar
product of the two states. Taking this into account, we can immediately
write down  $P_{I}$ for the optimal unambiguous phase-covariant multimeter,
\begin{equation}
P_{I}=\frac{1}{2^{N+1}}\sum_{k=1}^N \, C_{N+1,k}
\left|\langle\varphi_{N,k}^{+} |\varphi_{N,k}^{-}\rangle\right|
+\frac{1}{2^N}.
\label{pcPI}
\end{equation}
The contribution $2^{-N}$ to $P_{I}$ stems from the term $X$ that
is common to both operators $R_{\pm}$. On inserting the expression
(\ref{pcvarphiNk}) into Eq.~(\ref{pcPI}) we obtain
\begin{equation}
P_{I}=\frac{1}{2^{N+1}}\sum_{k=1}^N \,
\left| C_{N,k}-C_{N,k-1}\right| +\frac{1}{2^{N}}.
\label{pcPIsum}
\end{equation}
We must distinguish the cases of odd and even $N$. Let us assume
that $N$ is even ($N=2n$). We divide the sum in Eq.
(\ref{pcPIsum}) into two parts $k\leq N/2$ and $k>N/2$ and we find
\begin{eqnarray}
\sum_{k=1}^N
\left| C_{N,k} - C_{N,k-1}\right|= 2{N \choose N/2} -2.
\end{eqnarray}
On inserting the sum back into Eq.~(\ref{pcPIsum}) we obtain
\begin{equation}
P_{I}(2n)=\frac{1}{2^{2n}} {2n \choose n}. \label{pcPIoptimaleven}
\end{equation}
The calculation for odd $N=2n-1$ proceeds along similar lines and
one obtains
\begin{equation}
P_I(2n-1)=\frac{1}{2^{2n-1}}{2n-1 \choose n-1}.
\label{pcPIoptimalodd}
\end{equation}
It holds that $P_{I}(2n-1)=P_I(2n)$ hence the error-free
probabilistic phase-covariant multimeter with $2n-1$ qubit program
is exactly as efficient as the multimeter with $2n$-qubit program.
It is worth noting here that a similar behavior has been observed
in the context of optimal $1\rightarrow N$ phase covariant cloning
of qubits \cite{DArMa} where it was found that the global
fidelities of clones produced by the  $1\rightarrow 2n$ and
$1\rightarrow 2n+1$ cloning machines are equal. The asymptotic
behavior of the probability of inconclusive results
(\ref{pcPIoptimaleven}) and (\ref{pcPIoptimalodd}) can be
extracted with the help of the Stirling's formula $N!\approx
\sqrt{2\pi N}\, N^N e^{-N}$. On inserting this approximation into
(\ref{pcPIoptimaleven}) we get $P_{I}(N)\approx 2/\sqrt{2\pi N}.$

The POVM elements that describe the optimal error-free multimeter
can be easily written down as the properly weighted convex
 sum of the POVM elements that describe
the optimal unambiguous discrimination of the states
$|\varphi_{N,k}^{+}\rangle$  and $|\varphi_{N,k}^{-}\rangle$,
\begin{eqnarray}
\Pi_{+}=\sum_{k=1}^{N} D_{N,k}^{-1} |\varphi_{\perp,N,k}^{-}\rangle \langle
\varphi_{\perp,N,k}^{-}|,
\nonumber \\
\Pi_{-}=\sum_{k=1}^{N} D_{N,k}^{-1} |\varphi_{\perp,N,k}^{+}\rangle \langle
\varphi_{\perp,N,k}^{+}|,
\label{pcunambiguousPOVMoptimal}
\end{eqnarray}
and $\Pi_{?}=\openone-\Pi_{+}-\Pi_{-}$.
Here $|\varphi_{\perp,N,k}^{\pm}\rangle$ denote states orthogonal to
$|\varphi_{N,k}^{\pm}\rangle$, respectively,
\begin{equation}
|\varphi_{\perp,N,k}^{\pm}\rangle=
\sqrt{B_{N,k}}\, |0\rangle_{\mathrm{d}}|N,k\rangle_{\mathrm{p}} \mp
\sqrt{1-B_{N,k}}\, |1\rangle_{\mathrm{d}}|N,k-1\rangle_{\mathrm{p}},
\end{equation}
and
\begin{equation}
D_{N,k}= \frac{2}{N+1} \max(k,N+1-k).
\end{equation}
The effective three-component POVM on the data register associated with
POVM (\ref{pcunambiguousPOVMoptimal}) reads
\begin{equation}
\pi_{\pm}=(1-P_{I}) |\psi_{\pm}\rangle\langle \psi_{\pm}|,
\qquad \pi_{?}=P_I \openone.
\label{runningoutoflabels}
\end{equation}
Note, that when performing a generalized measurement described by the POVM
(\ref{runningoutoflabels})
the statistics of the sub-ensemble of conclusive results would exactly agree
with the statistics obtained by von Neumann projective measurement in basis
$|\psi_{\pm}\rangle$, so the multimeter indeed exactly
probabilistically performs the required measurement on the data qubit.

\subsection{Multimeter with a fixed fraction of inconclusive results\label{PCM-I}}

The deterministic multimeters and the error-free probabilistic
multimeters discussed in the preceding subsections can be considered as
special limiting cases of a more general class of optimal multimeters
that yield an inconclusive result with probability
$P_{I}=\mathrm{Tr}[\Pi_{?}(R_{+}+R_{-})/2]$
and give the correct measurement outcome  with probability
$P_{S}\leq 1-P_{I}$  when  the data register is prepared in the basis state
$|\psi_{+}(\phi)\rangle$ or $|\psi_{-}(\phi)\rangle$ with equal a-priori
probability.  It is convenient to introduce the relative success
rate
\begin{equation}
P_{RS}=\frac{P_{S}}{1-P_I}
\end{equation}
which gives the fraction of correct outcomes in the sub-ensemble
of conclusive results. Note that $P_{RS}$ can be also interpreted as
the average fidelity of the probabilistic multimeter.
The optimal multimeter should achieve the  maximal
possible $P_S$ (hence also $P_{RS}$) for a given fixed probability of
inconclusive results $P_{I}$. This class of multimeters is described by
three-component POVM similarly as the unambiguous (error-free) multimeter.
Such multimeters in fact perform
the optimal discrimination of mixed quantum states $R_{+}$ and $R_{-}$
with a fixed fraction of inconclusive results. This general
quantum-state discrimination scenario has  been recently analyzed
in detail in Refs.~\cite{FiJe,El}, where it was shown that
the optimal POVM must satisfy the following set of extremal equations:
\begin{equation}
\left(\lambda - \frac{1}{2}R_{\pm}\right)\Pi_{\pm}=0,
\qquad \left(\lambda-a R_{?}\right)\Pi_{?}=0,
\label{pcextremaleq}
\end{equation}
and
\begin{equation}
\lambda - \frac{1}{2}R_{\pm}\geq 0, \qquad \lambda-a R_{?}\geq 0.
\label{pcextremalineq}
\end{equation}
Here $R_{?}=(R_{+}+R_{-})/2$ and $\lambda$ and $a$ are Lagrange
multipliers that account for the constraints
$\Pi_{+}+\Pi_{-}+\Pi_{?}=\openone$  and
\begin{equation}
\mathrm{Tr}[\Pi_{?}R_{?}]=P_{I}.
\label{PIdef}
\end{equation}
It follows from the structure of the extremal
Eqs.~(\ref{pcextremaleq}) and (\ref{pcextremalineq}) that the
problem of optimal discrimination of two mixed states $R_{\pm}$
with a fraction of inconclusive results $P_{I}$ is formally
equivalent to the maximization of success rate of the
deterministic  discrimination of three mixed states $R_{+}$,
$R_{-}$, and $R_{?}$ with a-priori probabilities
$p_{\pm}=1/[2(a+1)]$ and $p_{?}=a/(a+1)$.  Of course, this
equivalence straightforwardly extends to discrimination of $n$
mixed states.

In the present case, the key simplification stems from the observation that
the operators $R_{\pm}$ have a common block diagonal form, which
was already explored in construction of the optimal error-free phase-covariant
multimeter. Formally, we can write
\begin{equation}
R_{\pm}=\frac{1}{2^{N+1}}\bigoplus_{k=0}^{N+1} R_{\pm,k}, 
\end{equation}
where
\begin{eqnarray*}
R_{\pm,k}&=&C_{N+1,k}|\varphi_{N,k}^{\pm}\rangle \langle \varphi_{N,k}^{\pm}|,
 \qquad k=1,\ldots,N,  \\
R_{\pm,0}&=&|0\rangle_{\mathrm{d}}\langle 0| \otimes |N,0\rangle_{\mathrm{p}}\langle N,0|, \\
R_{\pm,N+1}&=&|1\rangle_{\mathrm{d}}\langle 1|\otimes |N,N\rangle_{\mathrm{p}}\langle N,N|.
\end{eqnarray*}
Accordingly, the total Hilbert space of the data and the program states
$\mathcal{H}=\mathcal{H}_{\mathrm{d}}\otimes\mathcal{H}_{\mathrm{p}}$ can be decomposed into
a direct sum of
$\mathcal{H}_k$, $\mathcal{H}=\oplus_{k=0}^{N+1} \mathcal{H}_k$.
The Hilbert spaces $\mathcal{H}_k$ are either two-dimensional (spanned
by $|0\rangle_{\mathrm{d}}|N,k\rangle_{\mathrm{p}}$ and $|1\rangle_{\mathrm{d}}|N,k-1\rangle_{\mathrm{p}}$)
or one-dimensional (spanned by $|0\rangle_{\mathrm{d}}|N,0\rangle_{\mathrm{p}}$ or
$|1\rangle_{\mathrm{d}}|N,N\rangle_{\mathrm{p}}$).
The optimal $\Pi_{+}$, $\Pi_{-}$, $\Pi_{?}$ and $\lambda$ also have a
block-diagonal structure
\begin{equation}
\Pi_{\pm}=\bigoplus_{k=0}^{N+1} \Pi_{\pm,k},
\qquad \Pi_{?}=\bigoplus_{k=0}^{N+1} \Pi_{?,k},
\qquad \lambda=\bigoplus_{k=0}^{N+1} \lambda_{k}.
\label{pcdirectsum}
\end{equation}
The extremal equations (\ref{pcextremaleq}) and (\ref{pcextremalineq})
split into $N+2$  equations
\begin{equation}
\left(\lambda_k - \frac{1}{2}R_{\pm,k}\right)\Pi_{\pm,k}=0,
\qquad \left(\lambda_k-a R_{?,k}\right)\Pi_{?,k}=0,
\label{pcextremaleqk}
\end{equation}
\begin{equation}
\lambda_k - \frac{1}{2}R_{\pm,k}\geq 0, \qquad \lambda_k-a R_{?,k}\geq 0.
\end{equation}
We thus have to determine the optimal POVM on each subspace
$\mathcal{H}_k$ and then merge the solutions according to
(\ref{pcdirectsum}). Due to the structure of the operators
$R_{\pm}$, the task reduces to the discrimination of two pure
non-orthogonal states $|\varphi_{N,k}^{\pm}\rangle$ with
inconclusive results, which was discussed in detail by Chefles and
Barnett \cite{ChBa} and also by Zhang \emph{et al.} \cite{Zhang99}.

Let us first consider the non-degenerate case $k=1,\ldots,N$.
We have to distinguish the cases $C_{N,k}\geq C_{N,k-1}$ (i.e. $k\leq [N/2]$)
and $C_{N,k}<C_{N,k-1}$ ($k>[N/2]$).  We will explicitly present the results
for  $k\leq [N/2]$. The formulas for $k>[N/2]$ are similar and can be obtained
by simple exchanges $C_{N,k}\leftrightarrow C_{N,k-1}$ and
$|0\rangle_{\mathrm{d}}|N,k\rangle_{\mathrm{p}}
\leftrightarrow|1\rangle_{\mathrm{d}}|N,k-1\rangle_{\mathrm{p}}$.
The optimal POVM on each subspace $\mathcal{H}_k$ can be written as follows
\begin{eqnarray}
\Pi_{+,k}&=&\frac{1}{2\sin^2\Phi_k}|\Phi_{N,k}^{+}\rangle\langle\Phi_{N,k}^{+}|,
\nonumber \\
\Pi_{-,k}&=&\frac{1}{2\sin^2\Phi_k}|\Phi_{N,k}^{-}\rangle\langle\Phi_{N,k}^{-}|,
\nonumber \\
\Pi_{?,k}&=&(1-\tan^{-2}\Phi_k)|0\rangle_{\mathrm{d}}\langle 0|
\otimes |N,k\rangle_{\mathrm{p}}\langle N,k|,
\label{pcgeneralPOVMoptimal}
\end{eqnarray}
where
\begin{equation}
|\Phi_{N,k}^{+}\rangle=\cos\Phi_k|0\rangle_{\mathrm{d}}|N,k\rangle_{\mathrm{p}}\pm
\sin\Phi_k|1\rangle_{\mathrm{d}} |N,k-1\rangle_{\mathrm{p}}.
\end{equation}
The angle $\Phi_k$ is a function of the Lagrange multiplier $a$.
This dependence can be determined by substituting the explicit form of the
optimal POVM (\ref{pcgeneralPOVMoptimal}) into the extremal
Eqs.~(\ref{pcextremaleqk}) and solving the resulting system of
linear equations for $\lambda_k$ and $a$. After a bit tedious but otherwise
straightforward algebra we obtain
\begin{equation}
\tan\Phi_k=\left\{
\begin{array}{lll}
1, & & a<a_{\mathrm{th,k}},  \\
\sqrt{\frac{C_{N,k}}{C_{N,k-1}}}(2a-1), & & a \geq a_{\mathrm{th},k},
\end{array}
\right.
\label{pcPhik}
\end{equation}
where $a_{\mathrm{th},k}=\frac{1}{2}\left(1+\sqrt{C_{N,k-1}/C_{N,k}}\right)$.
The probability of inconclusive results $P_{I,k}$
and the probability of correct guess $P_{S,k}$
when discriminating the states $|\varphi_{N,k}^{\pm}\rangle$ with the POVM
(\ref{pcgeneralPOVMoptimal}) are given by
\begin{eqnarray}
P_{I,k}&=&\frac{C_{N,k}}{C_{N+1,k}}\left(1-\frac{1}{\tan^{2}\Phi_k}\right),
\nonumber \\
P_{S,k}&=&\frac{\cos^2(\Phi_k-\Theta_k)}{2\sin^2\Phi_k},
\label{pcPIPSoptimalk}
\end{eqnarray}
where $\Theta_k=\arctan(\sqrt{C_{N,k-1}/C_{N,k}})$.

The cases $k=0$ and $k=N+1$ require special treatment because the two states
to be discriminated are actually identical. Let us consider the case $k=0$.
If $a\neq 1/2$ then the optimal POVM can be formally determined from
Eqs.~(\ref{pcgeneralPOVMoptimal}) and  (\ref{pcPhik})
where the limit $C_{N,k-1}\rightarrow 0$ must be considered. One finds that
$\Pi_{?,0}=0$ for $a<1/2$ while $\Pi_{+,0}=\Pi_{-,0}=0$ and
$\Pi_{?,0}=\openone_0$ for $a>1/2$.
A sharp transition occurs at $a=1/2$ where the optimal POVM changes from
projective measurement  to a single-component POVM with all measurement
outcomes being interpreted as inconclusive results. The transition at $a=1/2$
can be described by a single parameter $\eta\in[0,1]$ and we can write
\begin{eqnarray*}
\Pi_{\pm,0}&=&\frac{1}{2}(1-\eta)|0\rangle_{\mathrm{d}}\langle 0|
\otimes |N,0\rangle_{\mathrm{p}}\langle N,0|,
\nonumber \\
\Pi_{?,0}&=&\eta|0\rangle_{\mathrm{d}}\langle 0| \otimes |N,0\rangle_{\mathrm{p}}\langle N,0|.
\end{eqnarray*}
Consequently, we have $P_{S,0}=1/2$, $P_{I,0}=0$ for $a<1/2$; $P_{S,0}=0$,
$P_{I,0}=1$ for $a>1/2$ and a smooth transition $P_{S,0}=(1-\eta)/2$,
$P_{I,0}=\eta$ at $a=1/2.$

\begin{figure}[!t!]
\centerline{\resizebox{0.9\hsize}{!}{\includegraphics*{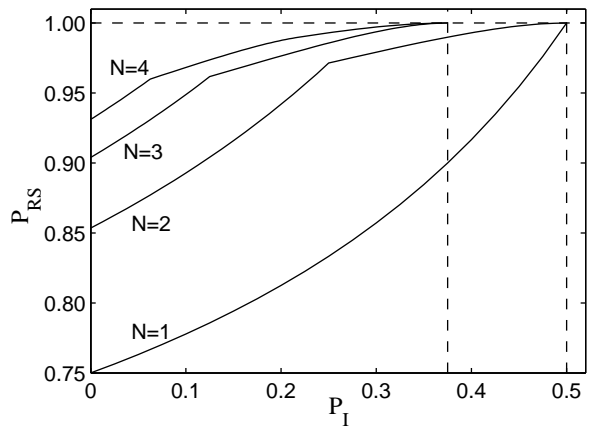}}}
\caption{Dependence of the relative success rate $P_{RS}$ of the optimal
phase-covariant multimeter with program $|\psi_+\rangle^{\otimes N}$ on the
fraction of inconclusive results $P_I$.}
\label{figPCM}
\end{figure}

The class of the optimal probabilistic phase covariant multimeters is
thus parameterized by two numbers $a\in[0,1]$ and $\eta\in[0,1]$.
If we combine all the above derived results we can express the
dependence of $P_S$  on $a$ and $\eta$ as follows,
\begin{equation}
P_{S}=\frac{1}{2^{N+1}}\sum_{k=1}^N C_{N+1,k} P_{S,k}
+ \frac{1}{2^{N+1}}(P_{S,0}+P_{S,N+1}),
\end{equation}
and a similar formula holds also for $P_I$. Rather than plotting
the dependence of $P_S$ and $P_I$ on $a$ and $\eta$, we directly
show in Fig.~\ref{figPCM} the dependence of the relative success
rate $P_{RS}=P_S/(1-P_I)$ (i.e., the fidelity of the probabilistic multimeter)
on the fraction of inconclusive results $P_I$.
We can see that $P_{RS}$  monotonically grows with $P_I$
and the point of unambiguous probabilistic operation is indicated
by $P_{RS}=1$, when $P_I$ has the value given by
Eqs.~(\ref{pcPIoptimaleven}) and (\ref{pcPIoptimalodd}).
Taking into account the symmetry of the POVM (\ref{pcgeneralPOVMoptimal})
with respect to the exchanges $k\rightarrow N-k+1$ and
$|0\rangle_{\mathrm{d}}\rightarrow
|1\rangle_\mathrm{d}$ it is easy to show that the effective POVM on the
data qubit corresponding to the optimal POVM (\ref{pcgeneralPOVMoptimal})
is given by
\begin{eqnarray*}
\pi_{\pm}&=&(1-P_{I})\left[P_{RS}|\psi_{\pm}\rangle\langle
\psi_{\pm}|+(1-P_{RS})|\psi_{\mp}\rangle\langle \psi_{\mp}| \right],
\nonumber \\
\pi_{?}&=&P_{I}\openone.
\end{eqnarray*}
The POVM has this structure for all possible  program states (i.e., all
measurement bases) hence the multimeter is indeed universal and covariant.
Since the POVM element $\pi_{?}$ is proportional to the identity operator
the detection of an inconclusive result does not provide any information
on the  data state.

\section{Universal multimeters for qubits\label{UM}}

In this section we will relax the confinement on the bases
consisting of vectors from the equator of the Bloch sphere and
will study universal multimeters designed for measurement in
\emph{any}  basis represented by two orthogonal states
$|\psi_{+}\rangle=\cos\frac{\vartheta}{2}|0\rangle
+e^{i\phi}\sin\frac{\vartheta}{2}|1\rangle$ and
$|\psi_{-}\rangle=\sin\frac{\vartheta}{2}|0\rangle
-e^{i\phi}\cos\frac{\vartheta}{2}|1\rangle$. We want this
measurement basis be controlled by the quantum state of a program
register, $|\Psi(\psi)\rangle_{\mathrm{p}}$. The program will be
assumed in the simplest symmetric form defining the measurement
basis: $|\Psi(\psi)\rangle_{\mathrm{p}} = |\psi_{+}\rangle|\psi_{-}\rangle$.

\subsection{Deterministic multimeter}

First, let us assume the multimeter that always ``works'' but that
allows for some erroneous results. Such a deterministic multimeter was
analyzed in Ref.~\cite{FiDuFi}. The optimal (in the sense of the
minimum error rate) two-component POVM  can be obtained
in the similar way as in Sec.~\ref{PCM-D}. In fact the task is
equivalent to the discrimination of two mixed states
\begin{eqnarray}
R_{+} &=& \int_\psi d\psi \, |\Psi_{+}\rangle\langle\Psi_{+}|,
\nonumber \\
R_{-} &=& \int_\psi d\psi \,
|\Psi_{-}\rangle\langle\Psi_{-}|, \label{stavy}
\end{eqnarray}
where averaging goes over all bases in the qubit space, i.e.,
over the whole surface of the Bloch sphere,
$\int_{\psi}d\psi= \frac{1}{4\pi}\int_0^\pi \int_0^{2\pi} \,
\sin\vartheta \,d\vartheta\, d\phi $, and
\begin{eqnarray*}
|\Psi_{+}\rangle &=& |\psi_{+}\rangle_{\mathrm{d}} \otimes
|\psi_+\rangle|\psi_{-}\rangle_{\mathrm{p}}, \\
|\Psi_{-}\rangle &=& |\psi_{-}\rangle_{\mathrm{d}} \otimes
|\psi_+\rangle|\psi_{-}\rangle_{\mathrm{p}}.
\end{eqnarray*}
After some algebra we obtain
\begin{equation}
R_{\pm}= \frac{1}{12} \Pi_{\mathrm{sym}}+\frac{1}{3} |A_{\pm}\rangle\langle
A_{\pm}|+\frac{1}{3} |B_{\pm}\rangle\langle B_{\pm}|,
\label{spectral}
\end{equation}
where $\Pi_{\mathrm{sym}}$ is the projector on the symmetric subspace
of three qubits and the eigenvectors $|A_{\pm}\rangle$ and $|B_{\pm}\rangle$
can be expressed in the computational basis as follows
\begin{eqnarray}
|A_{+}\rangle&=&\frac{1}{\sqrt{6}}(
|0\rangle_{\mathrm{d}} |11\rangle_{\mathrm{p}} +
|1\rangle_{\mathrm{d}} |01\rangle_{\mathrm{p}} -2 |1\rangle_{\mathrm{d}} |10\rangle_{\mathrm{p}}),
  \nonumber \\
|B_{+}\rangle&=&\frac{1}{\sqrt{6}}(
-2 |0\rangle_{\mathrm{d}} |01\rangle_{\mathrm{p}} + |0\rangle_{\mathrm{d}} |10\rangle_{\mathrm{p}}
+|1\rangle_{\mathrm{d}} |00\rangle_{\mathrm{p}});
  \nonumber \\[2mm]
|A_{-}\rangle&=&\frac{1}{\sqrt{6}}(
-|0\rangle_{\mathrm{d}} |11\rangle_{\mathrm{p}} + 2 |1\rangle_{\mathrm{d}} |01\rangle_{\mathrm{p}}
-|1\rangle_{\mathrm{d}} |10\rangle_{\mathrm{p}}),
  \nonumber \\
|B_{-}\rangle&=&\frac{1}{\sqrt{6}}(
-|0\rangle_{\mathrm{d}} |01\rangle_{\mathrm{p}} + 2 |0\rangle_{\mathrm{d}} |10\rangle_{\mathrm{p}}
- |1\rangle_{\mathrm{d}} |00\rangle_{\mathrm{p}}).
\nonumber \\
  \label{AB}
\end{eqnarray}
Notice the important orthogonality properties
\begin{eqnarray*}
\langle A_{+} | B_{+} \rangle =
\langle A_{-} | B_{-} \rangle &=&
\langle A_{+} | B_{-} \rangle =
\langle A_{-} | B_{+} \rangle = 0,
\\
\langle A_{+} | A_{-} \rangle &=&
\langle B_{+} | B_{-} \rangle = \frac{1}{2}.
\end{eqnarray*}
Moreover, the  states (\ref{AB}) are also orthogonal to any state from the
symmetric subspace  of three qubits.

As shown in Ref.~\cite{FiDuFi}, the optimal POVM for the deterministic
discrimination of the mixed states (\ref{spectral}) has the following form:
\begin{eqnarray}
\Pi_{+} &=& \frac{1}{2} \Pi_{\mathrm{sym}}+|\phi_1\rangle\langle \phi_1|
+|\phi_2\rangle\langle \phi_2|,
\nonumber \\
\Pi_{-} &=& \openone- \Pi_{+},
\label{PPM}
\end{eqnarray}
where $\openone$ is an identity operator on Hilbert space of three
qubits,  and
\begin{eqnarray}
|\phi_1\rangle&=&\frac{1}{2\sqrt{3}}[ (\sqrt{3}+1)|0\rangle_{\rm
d} |01\rangle_{\rm p} -(\sqrt{3}-1)|0\rangle_{\rm d}
|10\rangle_{\rm p}
\nonumber \\
&&-2\,|1\rangle_{\rm d} |00\rangle_{\rm p}],
\nonumber \\
|\phi_{2}\rangle&=&\frac{1}{2\sqrt{3}}[ (\sqrt{3}+1)|1\rangle_{\rm
d} |10\rangle_{\rm p} - (\sqrt{3}-1)|1\rangle_{\rm d}
|01\rangle_{\rm p}
\nonumber \\
&&-2\,|0\rangle_{\rm d} |11\rangle_{\rm p}]. \label{FI}
\end{eqnarray}
Corresponding maximal success rate (probability of a correct
result) is
\[
P_{S,\mathrm{max}}=\frac{1}{2}\left(1 +
\frac{1}{\sqrt{3}}\right).
\]
For any program
$|\psi_+\rangle|\psi_{-}\rangle_{\mathrm{p}}$ the effective POVM
on the data qubit is given by Eq.~(\ref{pcdataPOVMoptimal}) hence
the multimeter is universal and works equally well for all bases.

\subsection{Probabilistic error-free multimeter}

Let us now deal with the situation when we want to avoid any
errors. So we are looking for such a three-component POVM
($\Pi_{+}, \Pi_{-}, \Pi_{?}$) acting on data and program together
that gives three results according to the following prescription:
\begin{center}
\begin{picture}(150,43)(0,0)
 \put(-14,27){$|\psi_{+}\rangle_{\mathrm{d}} \otimes
 |\psi_{+}\rangle|\psi_{-}\rangle_{\mathrm{p}}$}

 \put(69,29.5){\vector(4,1){27}}
 \put(69,29.5){\vector(4,-1){27}}

 \put(-14,9){$|\psi_{-}\rangle_{\mathrm{d}} \otimes
 |\psi_{+}\rangle|\psi_{-}\rangle_{\mathrm{p}}$}

 \put(69,11.5){\vector(4,1){27}}
 \put(69,11.5){\vector(4,-1){27}}

 \put(102,35){$+$}

 \put(102,17.5){\,? (Do not know)}

 \put(102,0){$-$}

\end{picture}
\end{center}
Similarly as in Sec.~\ref{PCM}, the mean probability of an inconclusive result
is defined by $P_I=\frac{1}{2}\mathrm{Tr}[\Pi_{?}(R_{+} + R_{-})]$
and $\Pi_{?}$ is the POVM component corresponding to an inconclusive result.

Our aim is to find POVM that never wrongly identifies states
$|\Psi_{+}\rangle$ and $|\Psi_{-}\rangle$ for any choice of basis
$|\psi_{\pm}\rangle$ and that, at the same
time, minimizes the probability of inconclusive result.
This problem is formally equivalent to the determination of the optimal
POVM for unambiguous discrimination of two mixed states $R_{+}$ and $R_{-}$.
It means that, similarly as in Sec.~\ref{PCM-EF},
we are looking for operators $\Pi_{+}, \Pi_{-}, \Pi_{?}$
minimizing $P_I$ under the  constraints (\ref{pcconstraints}),
where the relevant $R_{\pm}$ are defined by Eq.~(\ref{stavy}).

The optimal POVM for the unambiguous discrimination of these two
mixed states consists of the multiples of projectors onto the
kernels of $R_{+}$ and $R_{-}$ (and of the supplement to unity).
The outcome $\Pi_{+}$ can be invoked only by $R_{+}$, the outcome
$\Pi_{-}$ only by $R_{-}$. We get
\begin{eqnarray}
\Pi_{+} &=& \frac{2}{3} \left[ |\chi_1\rangle\langle \chi_1|
+|\chi_2\rangle\langle \chi_2| \right],
\nonumber \\
\Pi_{-} &=& \frac{2}{3} \left[ |\kappa_1\rangle\langle \kappa_1|
+|\kappa_2\rangle\langle \kappa_2| \right],
\nonumber \\
\Pi_{?} &=& \openone - \Pi_{+} - \Pi_{-},
\end{eqnarray}
where
\begin{eqnarray*}
|\chi_1\rangle&=&\frac{1}{\sqrt{2}}(
|0\rangle_{\mathrm{d}} |01\rangle_{\mathrm{p}} - |1\rangle_{\mathrm{d}} |00\rangle_{\mathrm{p}}),\\
|\chi_2\rangle&=&\frac{1}{\sqrt{2}}(
|0\rangle_{\mathrm{d}} |11\rangle_{\mathrm{p}} - |1\rangle_{\mathrm{d}} |10\rangle_{\mathrm{p}}),\\
|\kappa_1\rangle&=&\frac{1}{\sqrt{2}}(
|0\rangle_{\mathrm{d}} |10\rangle_{\mathrm{p}} - |1\rangle_{\mathrm{d}} |00\rangle_{\mathrm{p}}),\\
|\kappa_2\rangle&=&\frac{1}{\sqrt{2}}(
|0\rangle_{\mathrm{d}} |11\rangle_{\mathrm{p}} - |1\rangle_{\mathrm{d}} |01\rangle_{\mathrm{p}}).
\end{eqnarray*}
This POVM leads to the lowest probability of inconclusive result
that equals $2/3$.

The proof of optimality follows the same lines as in
Sec.~\ref{PCM-EF}. Due to the particular structure of operators
$R_{+}$ and $R_{-}$ the problem of their unambiguous
discrimination splits into independent problems of the unambiguous
discrimination of two \emph{pure} states. This can be most easily seen from
the spectral decomposition of $R_{+}$ and $R_{-}$, cf. Eq. (\ref{spectral}).
Each operator $R_{\pm}$ possesses a 2-dimensional kernel and
the matrix representations of $R_{+}$ and $R_{-}$ exhibit a common
block-diagonal structure.  The first block (associated with eigenvalue $1/12$)
corresponds to the 4-dimensional symmetric subspace of
three qubits. The second block (associated with eigenvalue $1/3$)
corresponds to 2-dimensional
spaces spanned by $\{|A_+\rangle,~|B_{+}\rangle\}$ and
$\{ |A_{-}\rangle,~|B_{-}\rangle\}$, respectively.
Clearly, our discrimination problem reduces to
the unambiguous discrimination of states $| A_{+} \rangle$, $|
A_{-} \rangle,$ and $| B_{+} \rangle$, $| B_{-} \rangle$,
respectively.

Thus the minimal overall probability of the inconclusive result is
$$
P_I = \frac{1}{3} \left( \left| \langle A_{+} | A_{-} \rangle
\right| + \left| \langle B_{+} | B_{-} \rangle \right| \right)
+ \frac{4}{12} = \frac{2}{3}.
$$
The term $4/12$ stems from the totally symmetric states that are
the same for both operators $R_{\pm}$.

\subsection{Multimeter with a fixed fraction of inconclusive results}

Now we relax the requirement of unambiguous (error-free)
operation. Thus our task is: For given probability of inconclusive
result minimize the error rate (i.e., maximize the success rate)
or vice versa. We have already seen the two limit cases: The
deterministic and the probabilistic error-free multimeters as
described above.

The optimal discrimination of two mixed states $R_{\pm}$ with a
fraction of inconclusive results $P_{I}$ is formally equivalent to
the maximization of success rate of the deterministic
discrimination of three mixed states $R_{+}$, $R_{-}$, and $R_{?}
= (R_{+} + R_{-})/2$ with a-priori probabilities
$p_{\pm}=1/[2(a+1)]$ and $p_{?}=a/(a+1)$, where $a\in[0,1]$ is a
certain Lagrange multiplier \cite{FiJe,El}. Again, we can
profitably use the specific structure of operators $R_{\pm}$
described in the preceding subsection. The method of calculation
is the same as in Sec.~\ref{PCM-I}.

Let us start with the discrimination of vectors from the symmetric
subspace (let $\{ | \xi_i \rangle \}_i$ be an orthonormal basis in
$\mathcal{H}_{\mathrm{sym}}$). Because vectors $| \xi_i \rangle$
are the same for both $R_{\pm}$ we simply try to discriminate
identical states. It was shown in Sec.~\ref{PCM-I} that for $a<1/2$ the POVM
component corresponding to the inconclusive result $\Pi_{?,i}=0$
and for $a>1/2$ contrariwise the conclusive-result components are
zero, $\Pi_{\pm,i}=0$. For the boundary value $a=1/2$ there is a
smooth transition:
\begin{eqnarray*}
\Pi_{\pm,i} &=& \frac{1}{2}(1-\eta) \, | \xi_i \rangle \langle \xi_i |,
\nonumber \\
\Pi_{?,i} &=& \eta \, | \xi_i \rangle \langle \xi_i |, \quad \eta\in[0,1].
\end{eqnarray*}
The success rates and inconclusive-result rates are drawn in
Table~I.

\begin{table}[t]
\caption{Success rate and probability of inconclusive results as functions of
$a$ when discriminating two identical states.}

\begin{tabular}{|l|c|c|c|}
\toprule
   & $a<1/2$ & $a=1/2$ & $a>1/2$ \\
  \hline
$P''_{S}$ & 1/2 & $(1-\eta)/2$ & 0 \\
  \hline
$P''_{I}$ & 0 & $\eta$ & 1 \\
\botrule
\end{tabular}
\end{table}

Now we can proceed to the discrimination (with a given
inconclusive-result fraction) of states $| A_{+} \rangle$ and $|
A_{-} \rangle$ defined by Eqs.~(\ref{AB}). For states $| B_{+}
\rangle$ and $| B_{-} \rangle$ the calculation is completely
analogous and the results for success and inconclusive-result
rates are the same. States $| A_{\pm} \rangle$ include the angle
$60^\circ$ and they can be expressed in the following way:
$$
| A_{\pm} \rangle = \frac{1}{2} \left( \sqrt{3} \, | \beta \rangle \pm
| \alpha \rangle \right),
$$
where
\begin{eqnarray*}
| \alpha \rangle &=& \frac{1}{\sqrt{6}} \left(
2\, |0\rangle_{\mathrm{d}} |11\rangle_{\mathrm{p}} - |1\rangle_{\mathrm{d}} |01\rangle_{\mathrm{p}}
- |1\rangle_{\mathrm{d}} |10\rangle_{\mathrm{p}} \right),
\\
| \beta \rangle &=& \frac{1}{\sqrt{2}} \left(
|1\rangle_{\mathrm{d}} |01\rangle_{\mathrm{p}} - |1\rangle_{\mathrm{d}} |10\rangle_{\mathrm{p}} \right).
\end{eqnarray*}
POVM for the optimal discrimination can be written as follows
\begin{eqnarray}
\Pi_{\pm,A} &=& \frac{1}{2\sin^2\Phi} \,
| \Xi_{\pm} \rangle \langle  \Xi_{\pm} |,
\nonumber \\
\Pi_{?,A} &=& \left( 1-\frac{1}{\tan^2 \Phi} \right)
| \beta \rangle \langle \beta |,
\label{AA-POVMopt}
\end{eqnarray}
where
\begin{equation}
  | \Xi_{\pm} \rangle = \cos \Phi \, | \beta \rangle \pm
  \sin \Phi \, | \alpha \rangle.
 \label{Xi}
\end{equation}
We can imagine this POVM in the following geometrical way: We
start with $\Phi=45^\circ$ so that states $| \Xi_{\pm} \rangle$
are orthogonal. This situation corresponds to the Helstrom
deterministic (but erroneous) discrimination. Then, increasing
$\Phi$, we close vectors $| \Xi_{\pm} \rangle$ together on the
Bloch sphere. Finally, we get to the situation when $| \Xi_{+}
\rangle$ is orthogonal to $| A_{-} \rangle$ and $| \Xi_{-}
\rangle$ is orthogonal to $| A_{+} \rangle$; $\Phi=60^\circ$. This
case corresponds to the unambiguous discrimination of states $|
A_{\pm} \rangle$.

Now one can easily calculate the probability of success:
\begin{equation}
  P'_{S} = \frac{1}{8} \left( \frac{\sqrt{3}}{\tan \Phi} + 1 \right)^2,
 \label{AA-PS}
\end{equation}
and the probability of inconclusive result:
\begin{equation}
  P'_{I} = \frac{3}{4} \left( 1 - \frac{1}{\tan^2 \Phi} \right).
 \label{AA-PI}
\end{equation}
It follows from the extremal equations that
$$
\tan\Phi=\left\{
 \begin{array}{lll}
  1 & \mbox{~for~} & a < \frac{1}{2}(1+\frac{1}{\sqrt{3}}),  \\
  \sqrt{3}(2a-1) & \mbox{~for~} & a \geq \frac{1}{2}(1+\frac{1}{\sqrt{3}}).
 \end{array}
\right.
$$

\begin{figure}[!t!]
\centerline{\resizebox{0.9\hsize}{!}{\includegraphics*{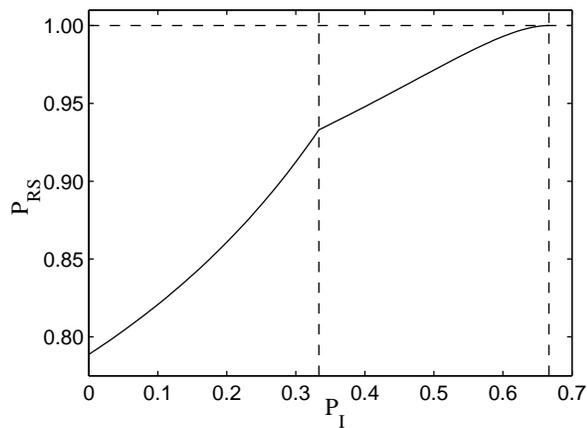}}}
 \caption{Dependence of the relative success rate, $P_{RS}$, on the probability
 of inconclusive results, $P_I$, for optimal universal multimeter with program
 $|\psi_{+}\rangle|\psi_{-}\rangle$.}
 \label{fig1}
\end{figure}

At this stage we are ready to write down the total success rate
and inconclusive-result rate for the discrimination of states $R_{\pm}$.
Clearly,
$$
P_S = \frac{1}{3} P''_S + \frac{2}{3} P'_S, \qquad
P_I = \frac{1}{3} P''_I + \frac{2}{3} P'_I.
$$
We can also introduce the relative success rate (i.e., the
success rate calculated only for ``conclusive'' results):
$ P_{RS} = {P_S}/{(1-P_I)} $.

One must examine four different sets of parameter $a$: $a \in
[0,\frac{1}{2})$, $a=\frac{1}{2}$, $a \in
(\frac{1}{2},\frac{1}{2}(1+\frac{1}{\sqrt{3}})]$,
and $a \in (\frac{1}{2}(1+\frac{1}{\sqrt{3}}),1]$.
Finally, it can be seen that (see also Fig.~\ref{fig1})
\begin{equation}
P_S = \left\{
\begin{array}{lll}
\displaystyle
\frac{1}{2} \left( 1 + \frac{1}{\sqrt{3}} \right) - \frac{P_I}{2}
 &\mbox{~~if~~}&
\displaystyle
0 \le P_I \le \frac{1}{3},
\\[4mm]
\displaystyle
\frac{1}{2} - \frac{P_I}{2} + \frac{1}{3} \sqrt{\frac{5}{4} - \frac{3 P_I}{2}}
 &\mbox{~~if~~}&
\displaystyle
\frac{1}{3} < P_I \le \frac{2}{3}.
\end{array} \right.
\label{UM-PS-PI}
\end{equation}
For $P_I=2/3$ the error-free operation ($P_{RS}=1$) is approached
and further increasing of $P_I$ has no reason.

Apparently, the optimal POVM for universal multimeters with fixed
fraction of inconclusive results has two different forms according
to the value of the probability of inconclusive result. First let
us write down the POVM for $P_I \in [0,\frac{1}{3}]$:
\begin{eqnarray}
\Pi_{\pm} &=& \Pi^{\mathrm{D}}_{\pm} - \frac{3}{2} P_I \, \Pi_{\mathrm{sym}},
\nonumber \\
\Pi_{?} &=& 3P_I \, \Pi_{\mathrm{sym}},
\label{povm1}
\end{eqnarray}
where $\Pi^{\mathrm{D}}_{\pm}$ denote the elements of POVM for
deterministic discrimination that are defined by Eq.~(\ref{PPM}).

When $P_I \in (\frac{1}{3},\frac{2}{3}]$ the POVM can be expressed
as
\begin{eqnarray}
\Pi_{\pm} &=& \Pi_{\pm,A} + \Pi_{\pm,B}, \nonumber \\
\Pi_{?} &=& \Pi_{\mathrm{sym}} + \Pi_{?,A} + \Pi_{?,B},
\label{povm2}
\end{eqnarray}
where $\Pi_{\pm,B}$ and $\Pi_{?,B}$ are POVM elements for
discrimination of vectors $| B_{\pm} \rangle$ that can be obtained
in a completely analogous way as that for vectors $| A_{\pm}
\rangle$ [see Eq.~(\ref{AA-POVMopt})]. For $P_I=\frac{1}{3}$ the
two POVM (\ref{povm1}) and (\ref{povm2}) coincide [notice, that
for $\Phi=45^\circ$, $|\Xi_+ \rangle = - |\phi_2 \rangle$ as
follows from Eqs.~(\ref{FI}) and (\ref{Xi})].

The operation of the multimeter for different values of $P_I$ can
be figured as follows: When $P_I$ grows from zero it is the most
convenient to gradually move the contributions, that are the same
for both $R_{\pm}$ and that substantially contribute to errors,
from conclusive to inconclusive results. It means the multiple of
the projector to the symmetric subspace increases in $\Pi_{?}$.
When $P_I = 1/3$ then $\Pi_{?} = \Pi_{\mathrm{sym}}$ and further
increase of the fraction of $\Pi_{\mathrm{sym}}$ is impossible
(because $\Pi_{\pm}$ and $\Pi_{?}$ must form a POVM). If one wants
to increase $P_I$ further above $1/3$ he/she must start to turn
vectors $| \Xi_{\pm} \rangle$ as described above. The point $P_I =
2/3$ corresponds to the unambiguous discrimination.

\section{Universal probabilistic error-free multimeter
         for qudits\label{qudits}}

Let us consider a multimeter that could realize an arbitrary von-Neumann
projective measurement on a single $d$-level system (qudit). Let
$|\psi_j\rangle$, $j=1,\ldots,d$ denote orthonormal-basis states.
We consider the
conceptually simplest program that consists of the $d$ qudits in basis states,
\begin{eqnarray*}
|\Psi\rangle_{\mathrm{p}}&=&
|\psi_1\rangle|\psi_2\rangle\ldots|\psi_j\rangle\ldots|\psi_{d}\rangle
\equiv [U_{d}(g)]^{\otimes d} |\Psi_0\rangle_{\mathrm{p}},
\end{eqnarray*}
where
$|\Psi_0\rangle_{\mathrm{p}}=|1\rangle|2\rangle\ldots|j\rangle\ldots|d\rangle$,
 $U_{d}(g)$ is a unitary operation acting on the basis states according to
$U_{d}(g)|j\rangle=|\psi_j\rangle$ and $g\in SU(d)$.
We are interested in the probabilistic error-free multimeter that
can respond with an inconclusive outcome but it never makes an error, i.e.
$\pi_j\propto |\psi_j\rangle\langle \psi_j|$.
The multimeter is described by a $(d+1)$-component POVM on $d+1$ qudits
(the data qudit  and $d$ program qudits). The POVM
$\{\Pi_1,\ldots,\Pi_{d},\Pi_{?}\}$ should optimally unambiguously discriminate
among $d$ mixed states
\begin{eqnarray}
R_{j}&=&\int_{SU(d)} U_{d}(g)|j\rangle_{\mathrm{data}} \langle j|U_{d}^{\dagger}(g)
 \nonumber \\
&&\otimes [U_{d}(g)]^{\otimes d}|\Psi_0\rangle_{\mathrm{p}}\langle \Psi_0|
[U_{d}^{\dagger}(g)]^{\otimes d} d\mu(g),
\end{eqnarray}
where the integration is carried over the whole group $SU(d)$ with the
invariant Haar measure $d\mu(g)$.

We conjecture that the optimal POVM elements $\Pi_j$ have the following
structure
\begin{eqnarray}
\Pi_j &=& C \, |\Sigma_{d}^{-}\rangle_{\bar{j}}\langle \Sigma_{d}^{-}|\otimes
\openone_j, \nonumber \\
\Pi_{?} &=& \openone - \sum_{j=1}^{d} \Pi_j,
\label{quditPOVM}
\end{eqnarray}
where $|\Sigma_{d}^{-}\rangle_{\bar{j}}$ is totally antisymmetric state of $d$ qudits:
the data qudit and all program qudits except for the $j$-th qudit, and
$\openone_j$ stands for the identity operator on the Hilbert space of the
$j$-th program qudit. We can write
\begin{equation}
|\Sigma_{d}^{-}\rangle_{\bar{j}}=\frac{1}{\sqrt{d!}}
\sum_{\bm{i}}\epsilon_{\bm{i}}|i_1\rangle_{\mathrm{data}}\otimes
|i_2,\ldots,i_{d}\rangle_{\mathrm{p} \bar{j}},
\label{quditSigma}
\end{equation}
where we sum over all permutations of $\{1,2,\ldots,d\}$ and
$\epsilon_{\bm{i}}$ is the sign of the permutation. Apparently,
vectors $|\Sigma_{d}^{-}\rangle_{\bar{j}} \otimes |x\rangle_{j}$,
where $|x\rangle_{j}$ is an arbitrary state of the $j$-th program qudit,
are orthogonal to any vector $|\Psi_k \rangle = |\psi_k
\rangle_{\mathrm{data}} |\psi_1 \rangle| \psi _2\rangle \dots
|\psi_{d} \rangle$ with $k \ne j$. It is easy to verify that
$\mathrm{Tr}[\Pi_j R_k]\propto \delta_{jk}$. It means, the only
contribution to the outcome $\Pi_j$ can originate from the $j$-th
basis state of the data qudit.

Clearly, POVM (\ref{quditPOVM}) is the POVM describing a
probabilistic unambiguous multimeter. We believe it is even the
optimal one. This hypothesis is based on the conjecture that
the kernels of operators $R_j$ have the form $\mathcal{K}_{j} =
\mathcal{H}^{\mathrm{ant}}_{j} \otimes \openone_{\bar{j}}$ where
$\mathcal{H}^{\mathrm{ant}}_{j}$ is the antisymmetric space of two
qudits --- the data one and $j$-th program one. Symbol
$\openone_{\bar{j}}$ denotes identity operator on $d-1$ program
qudits exclusive of $j$-th qudit. (At worst, $\mathcal{K}_{j}$ are
the subspaces of the appropriate kernels.) The $d$-dimensional
subspace spanned by $|\Sigma_{d}^{-}\rangle_{\bar{j}} \otimes
|x\rangle_{j}$, where $|x\rangle_{j}$ is an arbitrary state of the
$j$-th qudit, represents an intersection of $d-1$ spaces
$\mathcal{K}_{k}$ (excluding the $j$-th one):
$\bigcap_{\substack{k=1 \\ k \ne j}}^{d} \mathcal{K}_{k}$.

The sum of the $d$ POVM elements $\Pi_j$ must be lower than the identity
operator, $\sum_{j=1}^d \Pi_j \leq \openone,$
which imposes a constraint on the normalization factor $C$. Since we want to
maximize the probability of success we must choose the maximum possible $C$,
which can be expressed in terms of the maximum eigenvalue of the operator
\[
Y=\sum_{j=1}^{d} |\Sigma_{d}^{-}\rangle_{\bar{j}}\langle\Sigma_{d}^{-}|\otimes
\openone_{j}.
\]
The maximal admissible $C$ reads
\begin{equation}
C=\{\max[\mathrm{eig}(Y)]\}^{-1}.
\end{equation}
Instead of looking for the maximum eigenvalue of $Y$ we can equivalently
calculate the maximum eigenvalue of the operator
\begin{equation}
Z=\sum_{j=1}^d |f_j\rangle\langle f_j|,
\label{Z}
\end{equation}
where  $|f_j\rangle=|\Sigma_{d}^{-}\rangle_{\bar{j}}|1\rangle_j$.
The $d$ linearly independent states $|f_j\rangle$ span a
$d$-dimensional Hilbert space $\mathcal{H}_f$.  We can write
$|f_j\rangle=M|e_{j}\rangle$ where $|e_j\rangle$ form an
orthonormal basis in $\mathcal{H}_f$. On inserting this expression
into Eq.~(\ref{Z}) we find that
\begin{equation}
Z= \sum_{j=1}^d M|e_j\rangle\langle e_j| M^\dagger =MM^{\dagger},
\end{equation}
where the completeness of the basis $|e_j\rangle$ on
$\mathcal{H}_f$ has been used. It holds for any square matrix $M$
that $MM^\dagger$ has the same eigenvalues as $F=M^\dagger M$. In
the basis $|e_j\rangle$ the matrix elements of $F$ read
$F_{jk}=\langle f_j|f_k\rangle$. We thus have to determine the
scalar products of the non-orthogonal states $|f_j\rangle$.
Let us introduce unnormalized states of $d-1$ qudits,
$|\sigma_{d-1}^{-}\rangle_{\bar{j}\bar{k}}$, that are obtained by projecting
the $k$-th program qudit of the state $|\Sigma_{d}^{-}\rangle_{\bar{j}}$
onto state $|1\rangle_k$. It follows that $F_{jk}$ can be calculated
as a scalar product of  $|\sigma_{d-1}^{-}\rangle_{\bar{j}\bar{k}}$  and
$|\sigma_{d-1}^{-}\rangle_{\bar{k}\bar{j}}$,
\begin{equation}
F_{jk}=\,_{\bar{j}\bar{k}}\langle\sigma_{d-1}^{-}
|\sigma_{d-1}^{-}\rangle_{\bar{k}\bar{j}}
\label{quditFjkformula}
\end{equation}
It is easy to deduce from the Slater determinant representation of the
totally antisymmetric state (\ref{quditSigma}) that also
$|\sigma_{d-1}^{-}\rangle_{\bar{j}\bar{k}}$ is a totally antisymmetric state
of the data qudit and all the program qudits except $j$-th and $k$-th ones,
\begin{equation}
|\sigma_{d-1}^{-}\rangle_{\bar{j}\bar{k}}=\frac{(-1)^{t}}{\sqrt{d!}}\sum_{\bm{i}}
\epsilon'_{\bm{i}}|i_1\rangle_{\mathrm{data}}\otimes|i_2,\ldots,
i_{d-1}\rangle_{\mathrm{p},\bar{j}\bar{k}},
\label{quditsigmaminus}
\end{equation}
where one sums over all permutations of $\{2,3,\ldots,d\}$, $\epsilon'_{\bm{i}}$
is the sign of the permutation, and
\[
t=\left\{
\begin{array}{lll}
k   & \mbox{~for~} & j>k, \\
k-1 & \mbox{~for~} & j<k.
\end{array}
\right.
\]
Assuming $j\neq k$ and inserting the expressions
(\ref{quditsigmaminus}) into Eq.~(\ref{quditFjkformula}) we
immediately find  that
\begin{equation}
F_{jk}= \frac{(d-1)!}{d!} (-1)^{j+k-1}, \qquad j\neq k.
\end{equation}
Since $|f_j\rangle$ are normalized we finally have
\begin{equation}
F_{jk}=\delta_{jk}+(1-\delta_{jk})\frac{(-1)^{j+k-1}}{d}.
\end{equation}
The operator $F$ can be easily diagonalized,
\begin{equation}
F=\left(1+\frac{1}{d}\right)\openone-|\varphi_{d}\rangle\langle\varphi_{d}|,
\label{Ffinal}
\end{equation}
where $|\varphi_{d}\rangle=\frac{1}{\sqrt{d}}\sum_{j=1}^d(-1)^j|e_j\rangle$.
It follows from (\ref{Ffinal}) that the largest eigenvalue of $F$ is
$\mu_{\mathrm{max}}=1+1/d$ hence $C=d/(d+1)$ and the normalized POVM
(\ref{quditPOVM})
reads,
\begin{equation}
\Pi_j=\frac{d}{d+1} \,
|\Sigma_{d}^{-}\rangle_{\bar{j}}\langle \Sigma_{d}^{-}|\otimes \openone_j.
\end{equation}
By construction, the probabilistic multimeter is universal
and the  probability of success
\begin{equation}
 P_S =\frac{d}{(d+1)!}
 \end{equation}
does not depend on the particular basis chosen by the program state
and on the basis state $|\psi_j\rangle$ sent to the data register.
Consequently the multimeter indeed probabilistically
implements the projective measurement in the basis $\{ |\psi_j \rangle \}_{j=1}^d$
and the effective POVM on the data qudit reads
\begin{eqnarray}
\pi_j &=& P_S \, |\psi_j\rangle\langle \psi_j|, \qquad
j=1,\ldots,d,
\nonumber \\
\pi_{?} &=& (1-P_S) \openone.
\end{eqnarray}

\section{Conclusions\label{concl}}

In this paper we have investigated a broad class of quantum multimeters
that can perform a projective measurement on a single data qubit (or qudit).
The main feature of the quantum multimeters is that the measurement basis is
controlled by the quantum state of the program register that serves as a
kind of a quantum ``software'' while the multimeter itself (a quantum ``hardware'')
performs a fixed joint measurement on the data and program states.

In our investigations we have assumed finite-dimensional program register,
typically consisting of several qubits (or qudits). In this case it is
impossible to design perfect multimeter that would perform exactly and
deterministically the projective measurement in any basis from a continuous set,
with the basis being determined by the state of the program register.
The multimeters designed here are therefore only approximate.
Two conceptually different approximations have been considered. In the first
case, the multimeter operates deterministically and always produces
an outcome but the effective measurement on the data deviates from the ideal
projective measurement. Such errors are avoided in
the second approach when the multimeter is a probabilistic device whose
operation sometimes fails but, when it succeeds, then it carries out
exactly the desired projective measurement.

We have demonstrated that these two kinds of multimeters are in fact just limit
cases from a whole class of probabilistic multimeters that are characterized by
a certain fraction $P_I$ of the inconclusive results.  For a fixed dependence of
the program on the measurement basis,
the problem of designing the optimal multimeter  is formally equivalent to
finding the optimal POVM for discrimination of two mixed states. With the help
of the recently developed theory of optimal probabilistic discrimination of
mixed quantum states we have been able to analytically determine the optimal
phase-covariant multimeter for $N$-qubit program $|\psi_+\rangle^{\otimes N}$
as well as a universal multimeter with a two-qubit program
$|\psi_+\rangle|\psi_{-}\rangle$.  Remarkably, in both cases the success rate of
the optimal deterministic multimeter exactly coincides with the optimal fidelity
of estimation of the basis state $|\psi_{+}\rangle$ from a single copy of the
program state.

We have also proposed a generalization of the probabilistic error-free
multimeter to qudits assuming that the $d$-qudit program consists of
a product of the $d$ basis states. The construction of this multimeter
is inspired by  the structure of the optimal probabilistic multimeter for qubits
and relies on projections on totally anti-symmetric state of $d$ qudits.

Our findings clearly illustrate that the measurement on the data
qubit can be quite efficiently controlled by the quantum state of
the program register. In particular, we emphasize that a classical
description of the measurement basis would require infinitely many
bits of classical information, while only a few quantum bits
suffice in the present case to obtain an \emph{error-free}
(although probabilistic) operation. Our results also reveal many
intriguing connections between the concept of quantum multimeters,
discrimination of quantum states and optimal quantum state
estimation. This suggests that there might be also links to the
related problems of transmitting information about the direction
in space \cite{Bagan00,Peres01a,Bagan01a}  or about the reference
frame \cite{Peres01b,Bagan01b}  using the quantum states. For
instance, it is well known that the use of entangled states can
improve the fidelity of transmission in the two latter cases. A
natural question arises whether using entangled states as programs
one could achieve higher success rate (for a fixed size of the
program register) than with the product-state programs considered
in the present paper. More generally, one would ultimately like to
know what is the \emph{optimal program} leading to maximal
achievable success rate. This is a highly nontrivial optimization
problem that certainly deserves further investigation.

\begin{acknowledgments}
This research was supported under the projects LN00A015 and CEZ
J14/98 of the Ministry of Education of the Czech Republic.
JF also acknowledges support from the EU under projects RESQ
(IST-2001-37559) and CHIC (IST-2001-32150).
\end{acknowledgments}

\end{document}